# Interferometric Imaging of Nonlocal Electromechanical Power Transduction in Ferroelectric Domains

Lu Zheng[†,1], Hui Dong[†,2], Xiaoyu Wu[1], Yen-Lin Huang[1], Wenbo Wang[3], Weida Wu[3], Zheng Wang[*,2], Keji Lai[*,1]

[1] Department of Physics, University of Texas at Austin, Austin, TX 78712
[2] Department of Electrical and Computer Engineering, University of Texas at Austin, Austin, TX 78712
[3] Department of Physics and Astronomy, Rutgers University, Piscataway, NJ 08854
[†] These authors contributed equally to this work
[*] E-mails: zheng.wang@austin.utexas.edu; kejilai@physics.utexas.edu

## Abstract

**The electrical generation and detection of elastic waves are the foundation for acousto-electronic and acousto-optic systems. For surface-acoustic-wave devices, micro-/nano-electromechanical systems, and phononic crystals, tailoring the spatial variation of material properties such as piezoelectric and elastic tensors may bring significant improvements to the system performance. Due to the much smaller speed of sound than speed of light in solids, it is desirable to study various electroacoustic behaviors at the mesoscopic length scale. In this work, we demonstrate the interferometric imaging of electromechanical power transduction in ferroelectric lithium niobate domain structures by microwave impedance microscopy. In sharp contrast to the traditional standing-wave patterns caused by the superposition of counter-propagating waves, the constructive and destructive fringes in microwave dissipation images exhibit an intriguing one-wavelength periodicity. We show that such unusual interference patterns, which are fundamentally different from the acoustic displacement fields, stem from the nonlocal interaction between electric fields and elastic waves. The results are corroborated by numerical simulations taking into account the sign reversal of piezoelectric tensor in oppositely polarized domains. Our work paves new ways to probe nanoscale electroacoustic phenomena in complex structures by near-field electromagnetic imaging.**

The hallmark of wave interference, a ubiquitous phenomenon in nature, is the appearance of time-independent spatially varying patterns of the oscillation amplitude[1]. In the famous Young's double-slit experiment, alternating bright and dark bands on the detector screen vividly demonstrate the wave nature of light, where the periodicity of the interference pattern is proportional to the wavelength. Two counter-propagating waves, one usually generated by boundary-induced reflection of the other, can also interfere with each other to form a standing-wave pattern with a half-wavelength periodicity. In both cases, the interference fringes reveal the hidden phase information of the wave, which enables measurements with superior sensitivity, information capacity[2], and resolution[3] far beyond the wavelength limit. As a result, interferometry has become the basis for nearly all ultra-precision metrology in science and technology, ranging from astronomy[4] and quantum physics[5] to radar[6] and medical imaging[7].

Wave interference is generally caused by the superposition of local oscillating fields of individual waves at each point in the space. In this work, we demonstrate a special type of interference from the superposition of nonlocal interaction between electric fields and elastic waves in ferroelectric domain structures. Because of the sign reversal of piezoelectric tensor in oppositely polarized domains, the fringe patterns in microwave impedance maps are fundamentally different from that of the underlying acoustic fields. Our results are corroborated by first-principle numerical simulations. Microscopy on piezoelectric energy transduction is highly desirable for the design and characterization of novel surface acoustic wave (SAW) devices[8], microwave micro/nano-electromechanical systems (MEMS/NEMS)[9], and phonon-polariton systems[10]. In this context, our work may open a new research frontier to explore various nanoscale elastic phenomena in these systems by near-field electromagnetic imaging.

The experimental technique in this study is microwave impedance microscopy (MIM)[11], as schematically illustrated in Fig. 1A. The excitation signal $V = V_0 e^{i2\pi ft}$ (voltage $V_0$ around 0.1 V and frequency $f$ from 100 MHz to 10 GHz) is delivered to the center conductor of a shielded cantilever probe[12]. The tip can be viewed as a point voltage source since its diameter at the apex (~ 100 nm) is much smaller than the acoustic wavelength at these frequencies. The MIM electronics detect the real and imaginary components of the tip-sample admittance $Y = G + iB$ ($G$: conductance, $B$: susceptance), which are displayed as MIM-Re and MIM-Im images, respectively[13]. The MIM has been widely used for the study of nanoscale permittivity[14] and conductivity[15] distributions in complex systems. In the following, we will show that it can also reveal information on the electroacoustic power transduction in piezoelectric materials.

Our sample is single-crystalline lithium niobate ($LiNbO_3$), which is technologically important because of its high piezoelectric constants[16], low acoustic attenuation[17,18], and strong $2^{nd}$-order nonlinear optical coefficients[19,20]. $LiNbO_3$ has a trigonal (class 3m) crystal structure with a mirror *yz*-plane and a direct triad *z*-axis along the polar direction[16]. The polarization can be switched by electrical poling[19,21], allowing artificial domain patterns at micrometer sizes to be created for microwave signal processing[17,18] and nonlinear optics[19,20]. We start with the simplest scenario around a straight domain wall (DW) on a *z*-cut $LiNbO_3$ sample (Fig. 1A). At a first glance, the system is akin to the electron-wave interference near an atomic step edge imaged by scanning tunneling microscopy[22]. Since the domain inversion flips the sign of odd-rank tensors (the $1^{st}$ rank polarization *P* and the $3^{rd}$ rank piezoelectric tensor *e*)[16], the two oppositely polarized domains can be visualized by piezo-force microscopy (PFM) in Fig. 1B. The MIM data at $f = 967$ MHz in the same area are also displayed in Fig. 1B. The MIM-Im image, which represents the non-dissipative dielectric response, only shows weak contrast possibly due to the static surface charge. The MIM-Re image, on the other hand, exhibits clear interference fringes around the DW. Since the electrical conductance of $LiNbO_3$ due to free carriers is negligible, the MIM-Re contrast indicates that the microwave energy is dissipated through the piezoelectric transduction rather than the Ohmic loss.

Fig. 1C shows the averaged MIM-Re line profile across the DW. Neglecting a small spike on the wall due to the dielectric loss associated with DW vibrations[23], the main features include a prominent dip at the DW and damped oscillations with a periodicity of $\lambda$ away from the wall. Here $\lambda$ is found to be 4.55 µm from the Fourier transform of the ripples (inset of Fig. 1D) and the oscillation amplitude decays quadratically with the distance to the wall. We notice that the first pair of crests only develop as weak shoulders and the first pair of troughs are separated by 1.8 $\lambda$ rather than 2 $\lambda$. Similar MIM results are observed from 285 MHz to 6 GHz (Supplementary Fig. S1). As shown in Fig. 1D, the measured $1/\lambda$ scales linearly with *f* and the slope corresponds to an apparent phase velocity of $4.4 \pm 0.2$ km/s. Comparing it with the velocities of *x*-propagating acoustic waves[18,24,25] on *z*-cut $LiNbO_3$ (Table 1), it is clear that the results have the closest match to the pseudo surface acoustic wave (P-SAW)[26], whose dispersion lies in the continuum of bulk waves. Unlike the Rayleigh SAW (hereafter denoted as SAW) that exists on the surface of all solids[26], this electroacoustic Bleustein-Gulyaev[27,28] SAW only exists on the surface of piezoelectric materials. The displacement fields of the P-SAW are primarily polarized in the *y*-direction[29], although the wave is not purely transverse-horizontal due to the lack of an even-order symmetry axis in LiNbO$_3$.

It is tempting to interpret the MIM-Re fringes as the standing-wave patterns of the acoustic displacement fields underneath the tip, similar to those measured by scanning laser vibrometry[30,31], scanning electron microscopy[32,33], and scanning probe microscopy[34,35]. However, should the data represent a standing-wave pattern due to strong reflection off a DW, as suggested by an earlier MIM work[36], the measured periodicity would indicate the dominance of a guided wave with an extraordinarily large phase velocity of 8.8 km/s. In fact, since the acoustic impedance is the same for both domains, the reflection of displacement fields from the *yz*-DW is rather weak for both SAW and P-SAW, whereas the associated electric fields change sign across the DW due to the opposite piezoelectric coefficients (Supplementary Fig. S2). A careful analysis of the tip-sample interaction is therefore necessary to understand the intriguing interference pattern in Fig. 1B.

In LiNbO$_3$, the local mechanical strain and electric field are coupled by the piezoelectric effect. As a result, the tip displacement under an AC bias in the PFM measurement can be quantitatively analyzed by 3D finite-element modeling[37,38]. The MIM, on the other hand, measure the total power dissipation and numerical simulations have to take into account the energy transduction in the entire sample rather than the local displacement underneath the tip. Here, energy conservation dictates that the loss in electrical power $\vec{j} \cdot \vec{E}^*$ (*j*: current density; *E*: electric field) is equal to the mechanical power $\vec{F} \cdot \dot{\vec{u}}^*$ (*F*: electromechanical force density; $u$: displacement field; $\dot{u}$: time derivative of $u$), which excites various acoustic waves in solids[26]. The electromechanical force can be derived from the divergence of the stress field as $\vec{F} = \text{div } T$. Since P-SAW is dominated by the *y*-component of its displacement fields, the power transduction is predominantly determined by the overlap between $F_y$ and $\dot{u}_y$. Using vector calculus, one can show that $F_y$ is symmetric around the tip on a single domain (Supplementary Fig. S3). To satisfy the continuity condition, $\dot{u}_y$ is also an even function with respect to $x = x_{\text{tip}}$. Because of the sign flip of *e* across the DW, the overlap integral of $\text{Re}\left(\int F_y \cdot \dot{u}_y^* \, dx\right)$ in the two shaded areas in Fig. 2A cancels each other, leading to a drop in power transduction when the tip is close to the wall. As the tip moves away from the DW, the truncated overlap integral within $x \in (0, 2x_{\text{tip}})$ oscillates with the same periodicity as $\dot{u}_y$ and shows the power-law decrease of amplitude. Here the overlap integral of nonlocal acoustic wave sources are analogous to that in phase-array antennas[6], where distributed electromagnetic source configuration and position result in corresponding variations of the antenna impedance. Since the fringe patterns are a result of the interference of nonlocal power transduction rather than the local displacement fields of counter-propagating waves, oscillations in the microwave dissipation can be observed even in the absence of DW reflection. In reality, partial reflection of the SAW and P-SAW displacements always exist at the DW. The fact that only one-wavelength oscillation is visible in Fig. 1C, however, indicates that DW reflection is insignificant and does not change the overall picture here.

The qualitative picture above is confirmed by numerical simulations using finite-element analysis (FEA). Due to the prohibitive computational cost of a full 3D modeling for the sample volume exceeding 1000 $\lambda^3$, we truncate the material in the *y*-direction using periodic boundary condition, effectively treating the tip as an infinitely long line source (1 V at 1 GHz) along the *y*-axis (Supplementary Fig. S4). Except for the fact that the *E*-field decays as $1/r$ from the tip rather than $1/r^2$ in the actual case, this approximation captures the essential physics in our experiment, especially the phase velocities of the acoustic waves. Among the various acoustic waves excited by the tip (Fig. 2B), bulk waves do not contribute to periodic oscillations as the tip moves away from the DW. Further analysis (Supplementary Fig. S5) shows that the power of the P-SAW excited by the tip is ~ 5 times higher than that of the SAW, supporting our qualitative description above. As seen in Fig. 2C, the simulated power dissipation $\mathrm{Re}\left(\int \vec{j}^* \cdot \vec{E}\, d^3r\right)$ reproduces major features in the MIM-Re data, including the prominent dip at the DW and the damped oscillations with one-$\lambda$ periodicity. The use of a line source in the modeling is responsible for the $1/x$ rather than $1/x^2$ decay of the oscillation amplitude. The reduced separation between the first pair of troughs, 1.8 $\lambda$ rather than 2 $\lambda$, is also seen in the FEA result. We speculate that the deviation is due to DW reflection being no longer negligible when the tip-wall distance is less than one wavelength, although further work is needed to understand this behavior.

We now move on to the LiNbO$_3$ sample with two parallel DWs (PFM image in Fig. 3A). The corresponding MIM-Re images at 5 different frequencies are shown in Fig. 3B. The strength of the fringes increases as *f* goes from 850 MHz to 967 MHz and decreases as *f* further increases towards 1115 MHz. Figs. 3C and 3D display the simulated acoustic fields and microwave power loss at $f$ = 967 MHz. As summarized in Fig. 3E, when the DW spacing $d$ covers the distance between the first pair of troughs (~ 1.8 λ) and an integer multiple the wavelength, e.g., $d ≈ 1.8 λ + 2 λ$ at $f$ = 967 MHz, the two sets of ripples centered around the DWs reinforce each other, resulting in stronger peak-to-valley contrast. On the contrary, when $d – 1.8 λ ≈ 2 ± 0.5 λ$ at 850 MHz and 1115 MHz, the two sets of ripples are opposite in phase by 180°, resulting in suppressed oscillation strengths. More importantly, the ripples remain strong outside the two DWs at $f$ = 967 MHz, which is a direct evidence of the non-local power transduction nature of the MIM-Re images. If the data represent the standing-wave amplitude of the acoustic displacement fields[36], this on-resonance-like image should be free of oscillations outside the double-DW, since a Fabry-Perot interferometer does not reflect on resonance[1]. In addition, the fringes in Fig. 3B exhibits periodicity identical to the wavelength of the underlying acoustic wave, whereas the standing-wave patterns in a Fabry-Perot interferometer has a periodicity of half-λ.

Finally, we present the MIM results on several closed domain structures. Figs. 4A-D show the PFM and MIM-Re images (complete data in Supplementary Fig. S6) of four corral domains shaped in an equilateral triangle, a hexagon, a circle, and a square, respectively. Because of the crystal symmetry[16], DWs on the $z$-cut $LiNbO_3$ surface can only form straight lines along the three $y$-equivalent axes and become curved in other directions. Consequently, the domain designed to be a circle (Fig. 4C) appears as a rounded hexagon after electrical poling, and the domain designed to be a square (Fig. 4D) appears as a distorted rectangle. Beautiful interference patterns due to the superposition of ripples around each DW are observed in the MIM-Re images. For instance, the rectangular-lattice-like pattern in Fig. 4D can be viewed as an overlay of two sets of oscillations parallel to the $x$-axis and $y$-axis. The different velocities of P-SAW along the two directions[39], as calculated from the FEA (Fig. 4E), manifest in the different oscillation periods in the line profiles (Fig. 4F). Similar to the double-DW results, these features are different from the standing wave patterns in quantum corrals[40] in that the adjacent nodes are not spaced by half-λ. And the existence of such patterns does not indicate the presence of acoustic resonance[36]. In other words, the bright/dark regions in the MIM-Re images are not directly associated with the constructive/destructive interference of acoustic waves underneath the tip. Instead, they mark the tip locations around which the piezoelectric transduction over a distributed region of tens of microns, much wider than the closed domains themselves, is highly effective/ineffective.

Putting our findings in perspective, we have introduced a special type of interferometry by spatial mapping and numerical modeling of the electroacoustic power conversion in ferroelectric materials. The images of microwave dissipation reveal large internal degrees of freedom in piezoelectric and elastic tensors, which are not accessible by measurements of the acoustic displacement fields. For SAW devices, MEMS/NEMS, and phononic crystals, the spatial variation of piezoelectric effect can substantially influence the system performance. The submicron spatial resolution is also desirable to explore the effect of wave scattering, diffraction, and localization on the energy transduction. In all, microwave imaging may emerge as a powerful tool to probe the intimate coupling of electric and strain/stress fields in these systems.

## Methods:

**Sample preparation.** Congruent LiNbO$_3$ wafers from Gooch & Housego PLC (part number 99-00042-01) were used in this experiment. There was no intentional doping in the wafers to avoid additional composition-dependent wall structures. The wafers were poled to be a single domain before the fabrication. Standard photolithography was used to form the desired patterns on one side of the sample. A high voltage of 12 kV was applied across the 0.5 mm wafer in the patterned areas, resulting in polarization switching because the electric field exceeds the coercive field of 21 kV/mm. This electrical poling process was performed at the room temperature. The domain inversion flips the sign of odd-rank tensors, such as polarization $P$ (1$^{st}$ rank) and piezoelectric tensor $e$ (3$^{rd}$ rank), while leaving the even-rank tensors such as permittivity $\varepsilon$ (2$^{nd}$ rank) and elasticity $c$ (4$^{th}$ rank) unaltered. The single and double DWs in Figs. 1 and 3 are very long (> 1 mm) and well isolated (> 1 mm away) from other patterns. The corral domain structures in Fig. 4 are at least 50 μm away from the nearby patterns and the interaction between neighboring domains is small.

**Microwave impedance microscopy.** The MIM experiments were performed on an AFM platform (XE-70) from Park Systems. Details of the shielded MIM probe can be found in Ref. 12. Custom-built electronics were used for impedance imaging at frequencies ranging from 285 MHz to 6 GHz. Before the measurements, the phase of the reference signal to the mixer is adjusted such that the two channels are aligned to the real (MIM-Re) and imaginary (MIM-Im) parts of the admittance change, i.e., $V_{\text{MIM-Re}} \propto \Delta G \cdot V_0$ and $V_{\text{MIM-Im}} \propto \Delta B \cdot V_0$. The calibration of our 1 GHz MIM electronics shows that an output signal of 1 mV corresponds to an admittance contrast of 0.3 nS.

With a tip voltage $V_0 \sim 0.1$ V, the MIM-Re images can be interpreted by a simple conversion factor in that a $V_{MIM-Re}$ signal of 1 mV represents a transduced power $\frac{1}{2}\Delta G V_0^2$ of 1.5 pW.

**Numerical simulation.** The numerical simulation was performed by the Structural Mechanics Module in commercial finite-element analysis (FEA) software COMSOL 4.3. The acoustic fields in Figs. 2, 3, S2, S4 and the dissipated electrical power were simulated by the linear solver. The eigen-modes of SAW and P-SAW in Fig. 4 and Fig. S5 were calculated by the eigen-solver.

## Acknowledgements

We thank Z.-X. Shen, S.-W. Cheong, and S. Artyukhin for helpful discussions. The MIM work (L.Z., X.W., Y.-L.H., K.L.) was supported by NSF Division of Materials Research Award #1707372. The numerical simulation (H.D., Z.W.) was supported by the Packard Fellowships for Science and Engineering and NSF Division of Engineering Grant EFMA-1641069. The MIM instrumentation was supported by the U.S. Army Research Laboratory and the U.S. Army Research Office under grant W911NF1410483. W.Wang and W.Wu was supported by U.S. Department of Energy (DOE), Office of Science, Basic Energy Sciences (BES) under Award # DE-SC0018153.

## Author contributions

K.L. conceived and designed the experiments. W.Wang and W.Wu helped to design the samples. L.Z., X.W., Y.-L.H. performed the experiment. L.Z. and H.D. performed the numerical simulations. K.L. and Z.W. interpreted the data and wrote the paper. All authors contribute to data analysis and manuscript revision.

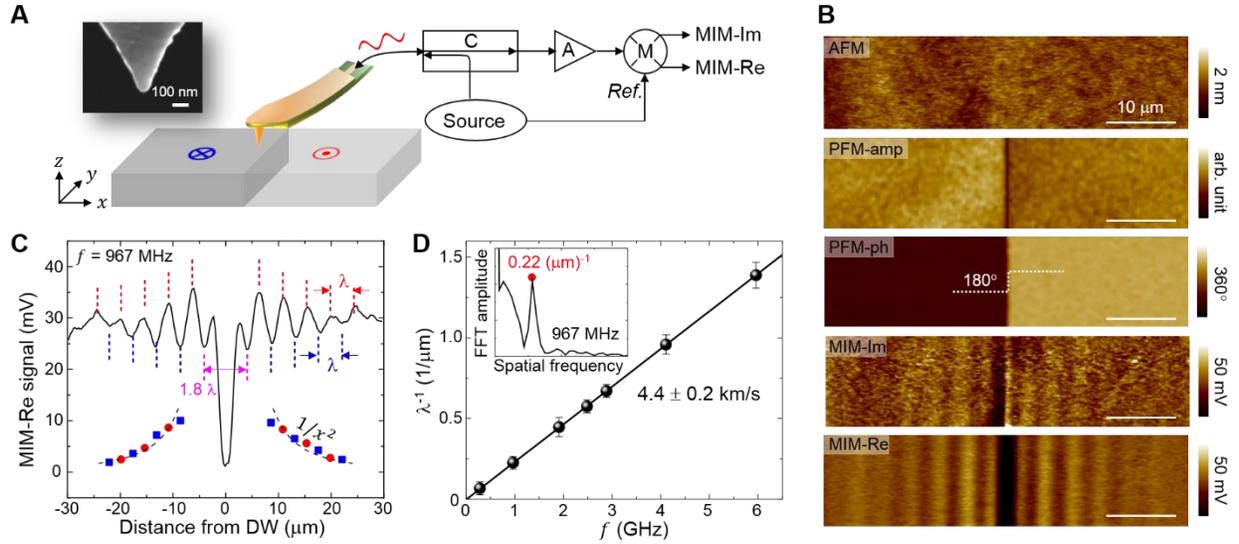

**Fig. 1.** Microwave imaging around a single LiNbO$_3$ domain wall. (*A*) Schematics of the MIM probe, electronics, and the *z*-cut LiNbO$_3$ sample with a single domain wall. The microwave signal is delivered to the cantilever tip by a directional coupler, and the reflected signal are amplified and mixed with the reference signal to form the MIM-Im and MIM-Re images. The top-right inset shows the scanning electron microscopy image of a typical tip apex. (*B*) From top to bottom: AFM, PFM amplitude (PFM-amp) and phase (PFM-ph) images, and MIM-Im/Re (*f* = 967 MHz) images of the sample. All scale bars are 10 μm. (*C*) MIM-Re line profile, in which the peaks and valleys are marked by red and blue dashed lines, respectively. Note that the first pair of troughs (labeled by pink dashed lines) are separated by 1.8 λ. The oscillation amplitude at each peak (red circles) and valley (blue squares) is the difference between its signal and the average signal of the two adjacent valleys and peaks, respectively. The black dashed lines are fits to the inverse square of the distance to the wall. (*D*) Linear relation between λ$^{-1}$ and the frequency. The slope corresponds to a wave velocity of 4.4 ± 0.2 km/s. The inset shows the Fourier transform of the data in **c** with a spatial frequency of 0.22 (μm)$^{-1}$.

| Annotation | Velocity (km/s) | Properties | Polarization |
|---|---|---|---|
| Rayleigh surface wave | 3.82 | Surface guided, non-leaky | $x$- and $z$-dominant |
| Slow transverse bulk wave | 4.08 | Bulk continuum | $y$-dominant |
| Bleustein-Gulyaev pseudo surface wave | 4.55 | Surface guided, leaky | $y$-dominant |
| Fast transverse bulk wave | 4.79 | Bulk continuum | $z$-dominant |
| Longitudinal bulk wave | 6.55 | Bulk continuum | $x$-dominant |

**Table 1.** Velocities and properties of the acoustic waves propagating along the $x$-axis on $z$-cut LiNbO$_3$ surface. The Rayleigh SAW is confined to the surface (non-leaky). The Bleustein-Gulyaev pseudo-SAW has a velocity greater than that of the slow transverse bulk wave. Its energy therefore leaks to the bulk during the propagation.

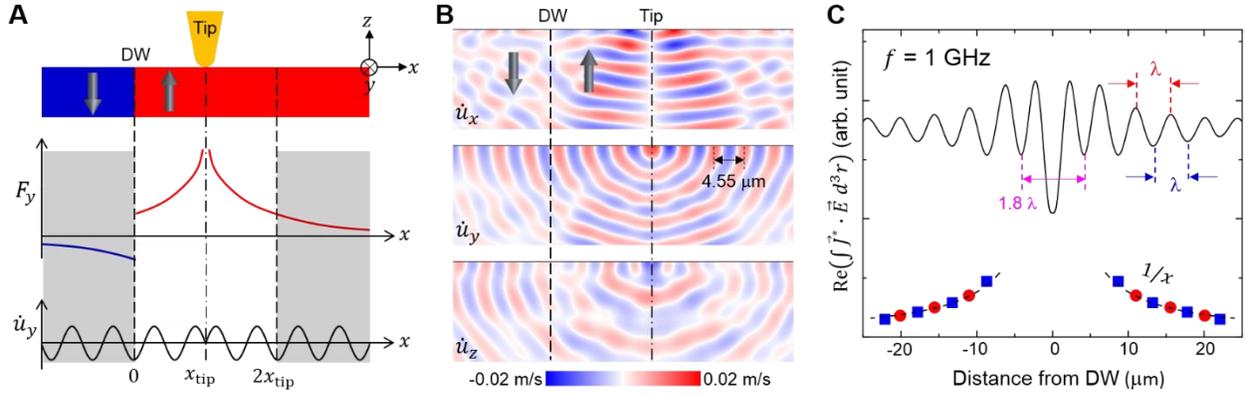

**Fig. 2.** Numerical simulation of the power transduction near a single domain wall. (*A*) (top) Schematic view of the tip-sample configuration in the *xz*-plane. (middle) Sketches of the *y*-components of the mechanical force and (bottom) time derivative of the displacement near the surface. The sign of $F_y$ is flipped in opposite domains due to the sign reversal of piezoelectric coefficients. The overlap integral of $F_y \cdot \dot{u}_y$ in the two shaded areas cancels each other. (*B*) (top to bottom) Snap shots of the *x*-, *y*-, and *z*-components of the simulated time derivative of displacement fields in the *xz*-plane. The P-SAW with a wavelength of 4.55 μm at 1 GHz is seen in the *y*-component of the velocity fields. (*C*) Numerical simulation of the dissipated electrical power as a function of the tip position. The strength of the oscillation amplitude scales with $1/x$ (dashed lines) instead of $1/x^2$ from the wall due to the use of a line source in the modeling.

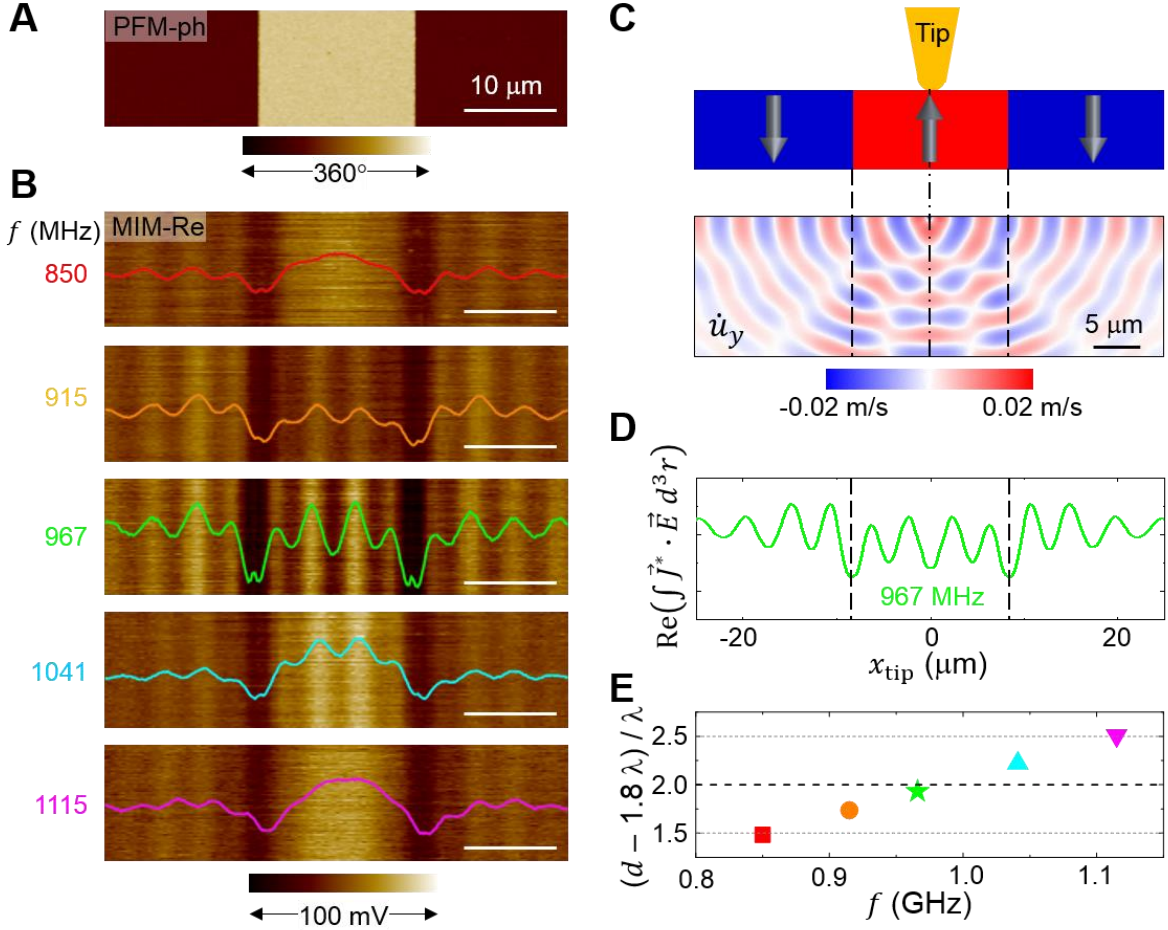

**Fig. 3.** Imaging and simulation of the double-DW sample. (*A*) Out-of-plane PFM phase image of the double-DW sample. (*B*) MIM-Re images and the corresponding line profiles at 5 different frequencies. All scale bars are 10 μm. (*C*) Tip-sample configuration (top) and simulated $\dot{u}_y$ field at 967 MHz (bottom) of the double-DW sample. (*D*) Simulated electrical power dissipation as a function of the tip position. The two dashed lines indicate the locations of the two walls. (*E*) $(d - 1.8\lambda)/\lambda$ as a function of $f$. The 5 frequency points, which are color coded as that in (*B*), are selected such that $(d - 1.8\lambda)$ roughly equals to 1.5 λ, 1.75 λ, 2.0 λ, 2.25 λ, and 2.5 λ, respectively.

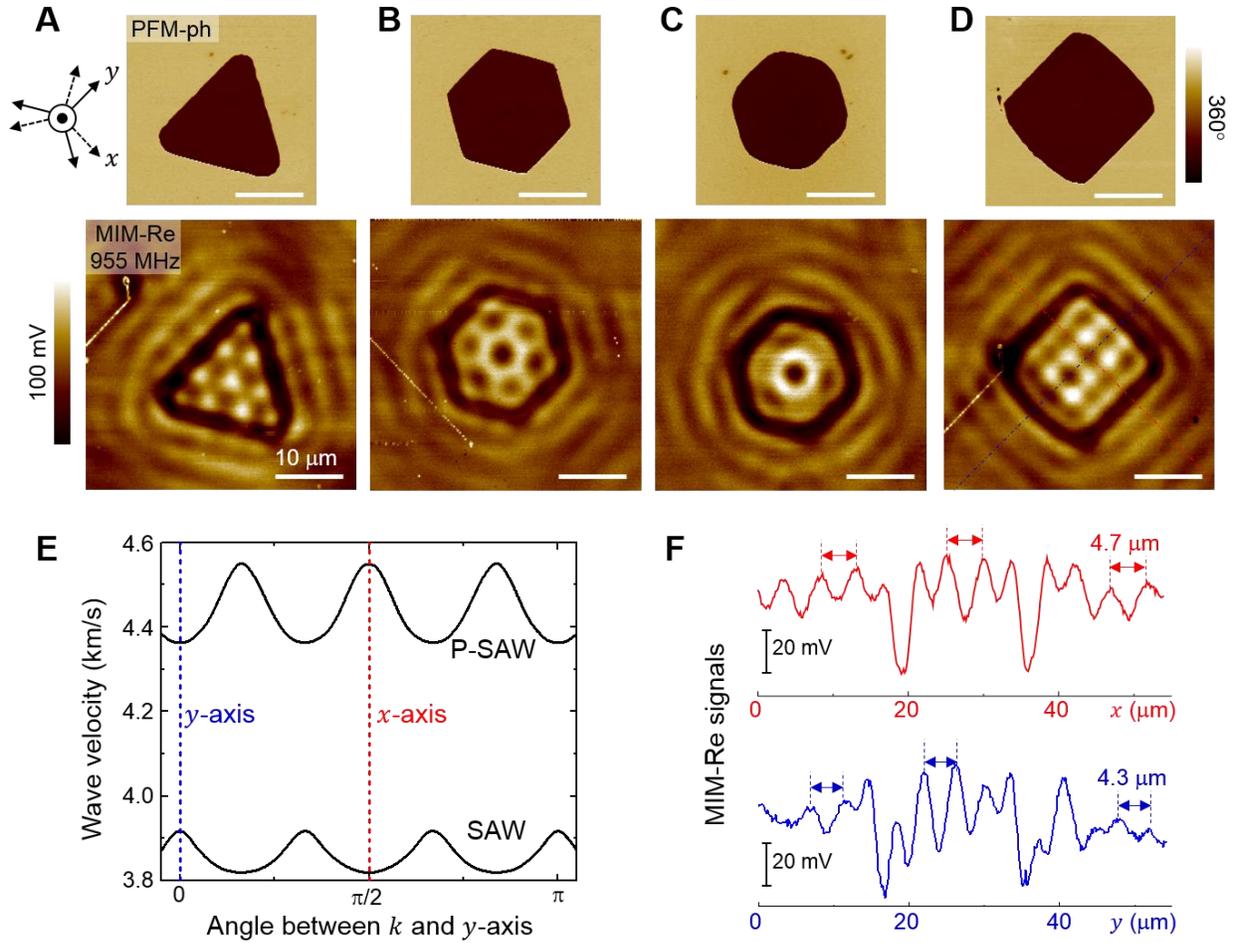

**Fig. 4.** Interference of piezoelectric transduction in corral domains. (*A-D*) PFM-ph (top) and MIM-Re images at $f$ = 955 MHz (bottom) of four closed LiNbO$_3$ domains. Clear interference patterns due to the superposition of ripples around each DW are seen in the MIM-Re data. All scale bars are 10 μm. (*E*) Velocities of P-SAW and SAW calculated from the COMSOL eigen-solver as a function of the angle between propagation vector $k$ and the y-axis on the z-cut LiNbO$_3$ surface. (*F*) Line profiles in (*D*), showing different oscillation periods along *x*-direction (red) and *y*-direction (blue). The result is consistent with the higher P-SAW velocity along the *x*-axis than that along the *y*-axis.

# Supplementary Information

# Interferometric Imaging of Nonlocal Electromechanical Power Transduction in Ferroelectric Domains

Lu Zheng[†,1], Hui Dong[†,2], Xiaoyu Wu[1], Yen-Lin Huang[1], Wenbo Wang[3], Weida Wu[3], Zheng Wang[*,2], Keji Lai[*,1]

[1] Department of Physics, University of Texas at Austin, Austin, TX 78712

[2] Department of Electrical and Computer Engineering, University of Texas at Austin, Austin, TX 78712

[3] Department of Physics and Astronomy, Rutgers University, Piscataway, NJ 08854

[†] These authors contributed equally to this work

[*] E-mails: zheng.wang@austin.utexas.edu ; kejilai@physics.utexas.edu

## Section S1: Full set of MIM data of the single DW

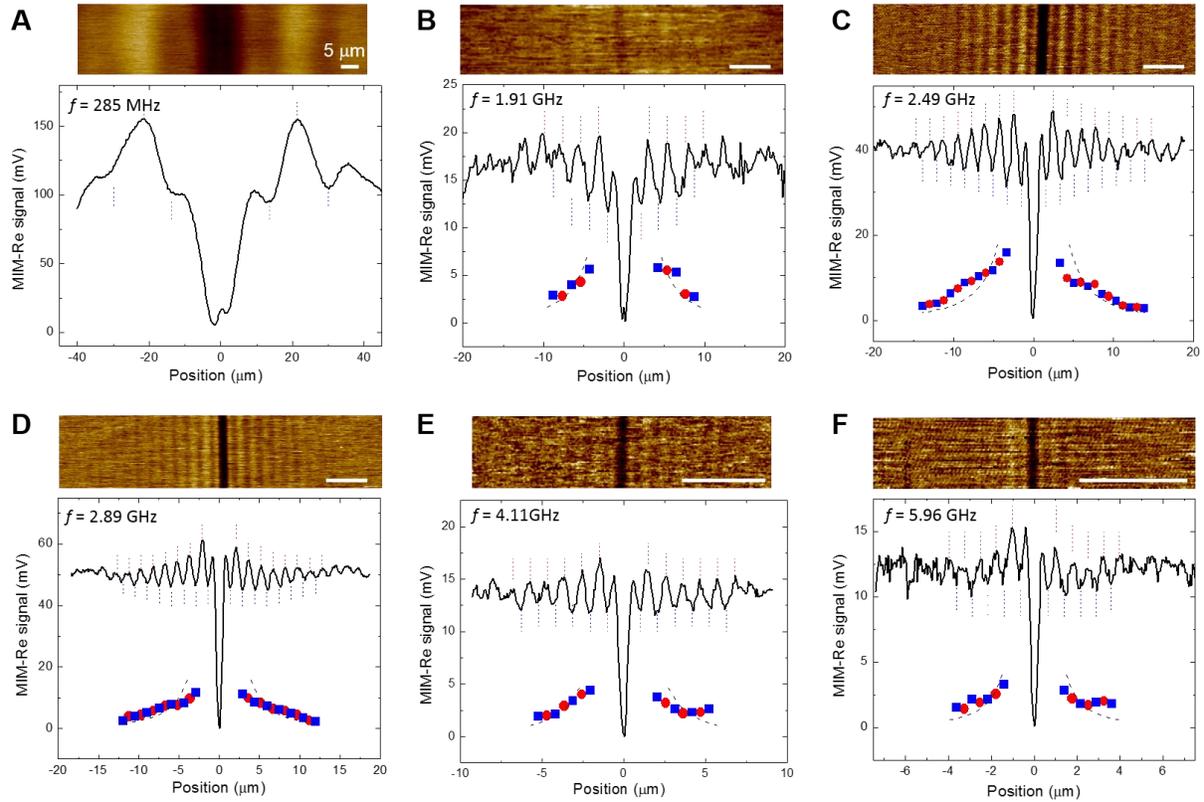

**Fig. S1.** MIM-Re images and the corresponding line profiles (averaged over 100 repeated line scans) at 6 different frequencies. The first two dips nearby the DW are labeled by pink dashed lines. Peaks and other valleys are labeled by red and blue dashed lines, respectively. The oscillation amplitude of each peak (valley) is determined as the difference between its signal and the average signal of nearby valleys (peaks). The black dashed lines are fits to the inverse square of the distance to the DW. All scale bars are 5 μm.

Fig. S1 shows the MIM-Re images at 6 different frequencies, from which the data points in Fig. 1D are extracted. To obtain the line profiles, we average over 100 repeated scans to improve the signal-to-noise ratio. The main features in Fig. 1C, such as the prominent dip centered at the DW and a series of damped oscillations away from the wall, are seen at frequencies ranging from 285 MHz to 5.96 GHz.

**Section S2: DW reflection of SAW and P-SAW**

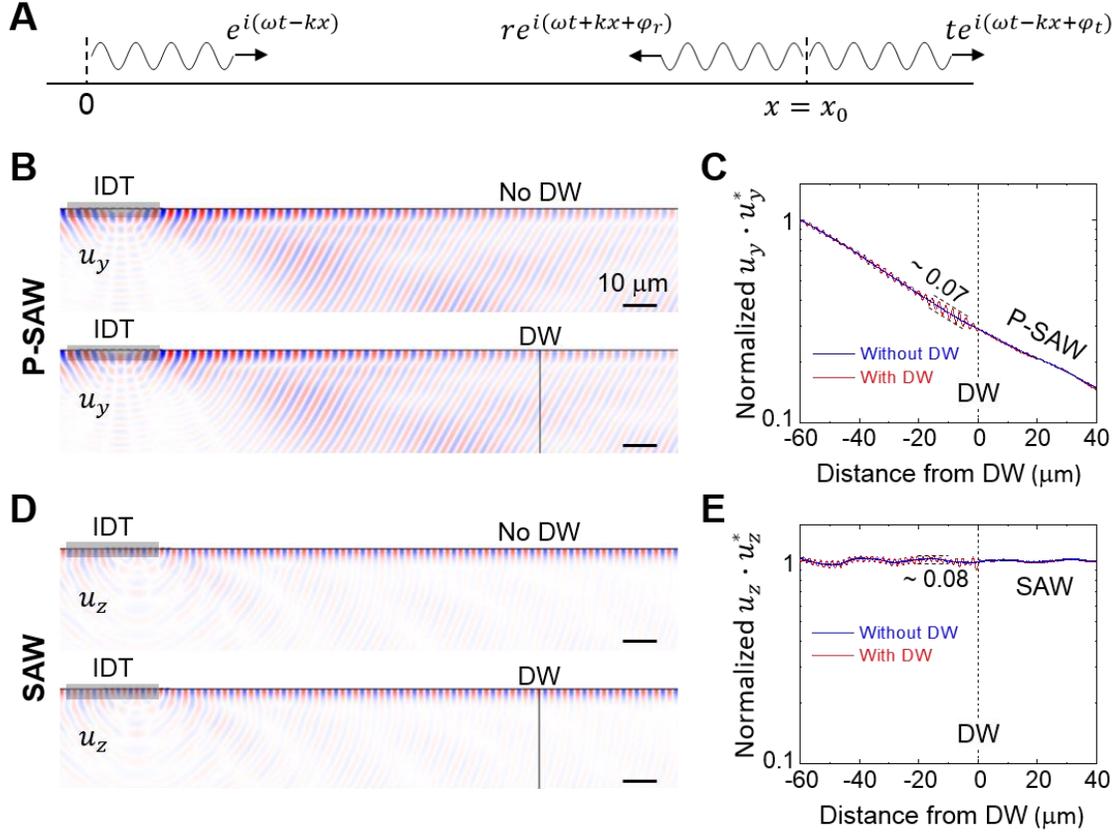

**Fig. S2**. (*A*) Schematic of the one-dimensional incident, reflected, and transmitted waves. (*B*) Simulated *y*-component of the displacement map without (top) and with the DW (bottom), where the spacing of the interdigital transducer (IDT) matches the wavelength of the P-SAW. (*C*) Normalized $u_y \cdot u_y^*$ near the surface as a function of the distance from the DW. (*D*) Simulated *z*-component of the displacement map without (top) and with the DW (bottom), where the spacing of the IDT matches that of the Rayleigh-like SAW. (*E*) Normalized $u_z \cdot u_z^*$ near the surface as a function of the distance from the DW.

The reflection of the displacement fields of Rayleigh-type SAW (denoted as SAW) [*S1*] and Bleustein-Gulyaev-type pseudo-SAW (denoted as P-SAW) [*S2*, *S3*] from the DW located in the *yz*-plane can be analyzed by FEA modeling. Assuming a +*x*-propagating wave (ω: frequency, *k*: wave vector, α: attenuation coefficient) is reflected at $x = x_0$ with a reflection coefficient of *r* and a phase slip of φ, the total displacement within $x \in (0, x_0)$ can be expressed as follows (Fig. S2A).

$$u = e^{i(\omega t - kx)} \cdot e^{-\alpha x} + r e^{i(\omega t + kx + \varphi)} \cdot e^{-\alpha(x_0 - x)} \tag{S2.1}$$

The attenuation and reflection coefficients can be evaluated by taking the product of *u* and its complex conjugate. The four situations based on the values of *r* and α are listed below.

$$u \cdot u^* = \begin{cases} 1 & [\alpha = 0, r = 0] \\ 1 + r^2 + 2r\cos(2kx + \varphi) & [\alpha = 0, r \neq 0] \\ e^{-2\alpha x} & [\alpha \neq 0, r = 0] \\ e^{-2\alpha x} + r^2 e^{-2\alpha(x_0 - x)} + 2re^{-\alpha x_0}\cos(2kx + \varphi) & [\alpha \neq 0, r \neq 0] \end{cases} \quad (S2.2)$$

Fig. S2B shows the simulated *y*-component of the displacement fields, where the source is an interdigital transducer (IDT) ~ 100 μm away from the DW. The spacing in the IDT is set to excite the P-SAW only. From the wave amplitude near the surface, it is obvious that the energy of the P-SAW continuously leaks into the bulk (slow transverse bulk wave) as it propagates away from the source. In Fig. S2C, we plot the simulated $u_y \cdot u_y^*$ at ~100 nm below the surface. By fitting to an exponential decay, we can obtain a wave attenuation of ~0.4 dB/wavelength, consistent with the literature [*S4*]. The simulated results with a DW are shown in Figs. S2B and S2C. The small modulation in $u_y \cdot u_y^*$ is due to the DW reflection. From Eq. (S2.2), one can show that $4re^{-\alpha x_0} = 0.07$ and the reflection coefficient $r \approx 2.6\%$. Similar results are also obtained for the Rayleigh-type SAW in Figs. S2D and S2E, which does not attenuate during the propagation. Here $u_z$ is analyzed since the SAW is primarily polarized along the *z*-direction. A reflection coefficient of $r \approx 2\%$ is calculated for the SAW. The voltage standing wave ratio (VSWR) can be calculated as follows.

$$\text{VSWR} = \frac{1+r}{1-r} \quad (S2.3)$$

Due to the small *r*, the VSWR ≈ 1.05 is very close to 1 for both SAW and P-SAW, indicative of the weak DW reflection for these surface waves. We emphasize that since the acoustic impedance is the same for up and down domains, a vanishing *r* is indeed expected for the displacement fields. The electric fields associated with the propagating waves, however, will experience a sign reversal across the DW due to the opposite piezoelectric coupling coefficient.

**Section S3. Electromechanical force density**

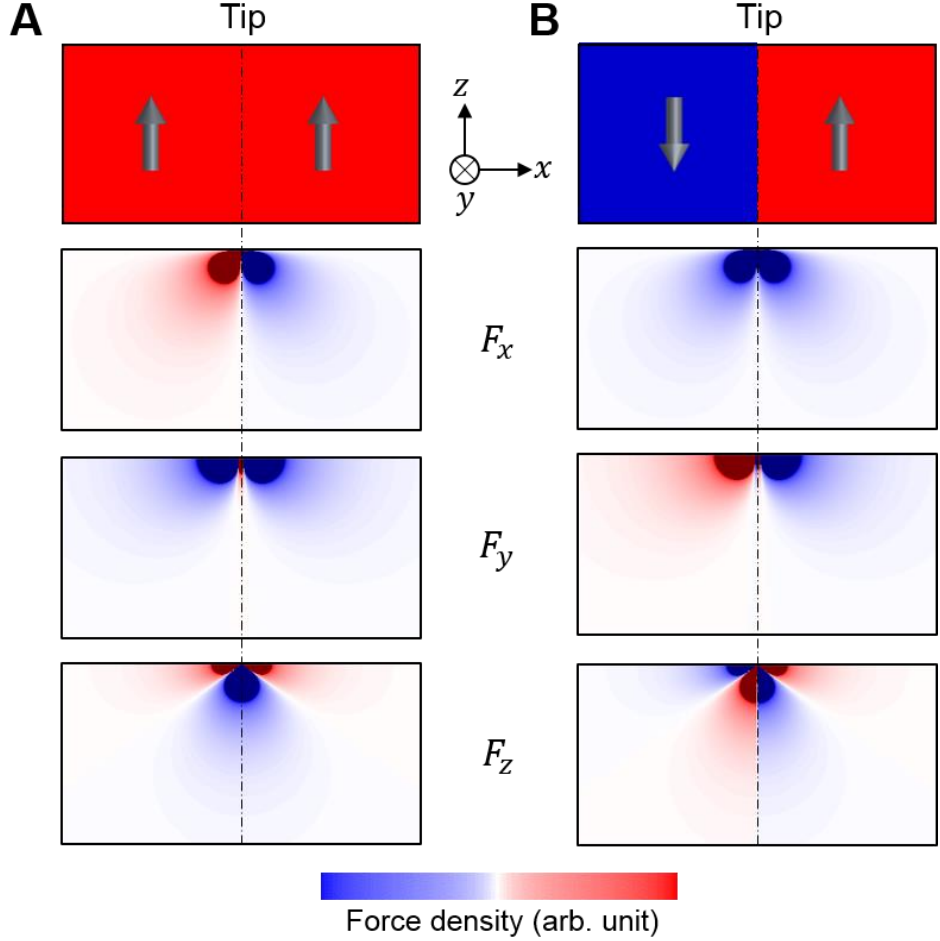

**Fig. S3**. Three components of the mechanical force density (force per unit volume) analytically calculated from the piezoelectric coupling when (*A*) the tip is on top of one single domain and (*B*) the tip is on the DW.

The MIM tip can be viewed as a point voltage source in the sagittal *xz*-plane since its diameter is much smaller than the acoustic wavelength at GHz frequencies. The three components of the electric field from such a point charge are as follows.

$$E_x = \frac{q}{4\pi\epsilon_0} \cdot \frac{x}{r^3}, \qquad E_y = \frac{q}{4\pi\epsilon_0} \cdot \frac{y}{r^3}, \qquad E_z = \frac{q}{4\pi\epsilon_0} \cdot \frac{z}{r^3} \qquad (S3.1)$$

The electromechanical force induced by the piezoelectric coupling can be calculated as the divergence of the stress tensor $T = cS - e^T E$ [*S5*]. For simplicity, we only consider the second term associated with the external electric fields.

$$\vec{F} = \text{div}\, T_{\text{ext}} = \nabla \cdot (-e^T E) \qquad (S3.2)$$

We can then plug the piezoelectric tensor of LiNbO$_3$ [*S6*, *S7*] into the equation as follows.

$$-\begin{bmatrix} e_{11} & e_{21} & e_{31} \\ e_{12} & e_{22} & e_{32} \\ e_{13} & e_{23} & e_{33} \\ e_{14} & e_{24} & e_{34} \\ e_{15} & e_{25} & e_{35} \\ e_{16} & e_{26} & e_{36} \end{bmatrix} \begin{bmatrix} E_x \\ E_y \\ E_z \end{bmatrix} = -\begin{bmatrix} 0 & -2.5 & 0.23 \\ 0 & 2.5 & 0.23 \\ 0 & 0 & 1.3 \\ 0 & 3.7 & 0 \\ 3.7 & 0 & 0 \\ -2.5 & 0 & 0 \end{bmatrix} \begin{bmatrix} E_x \\ E_y \\ E_z \end{bmatrix} \quad (S3.3)$$

Using vector calculus, the three components of the force density as the divergence of the external stress tensor are:

$$\begin{aligned} F_x &= 2.5 \partial_x E_y - 0.23 \partial_x E_z + 2.5 \partial_y E_x - 3.7 \partial_z E_x \\ F_y &= 2.5 \partial_x E_x - 2.5 \partial_y E_y - 0.23 \partial_y E_z - 3.7 \partial_z E_y \\ F_z &= -3.7 \partial_x E_x - 3.7 \partial_y E_y - 1.3 \partial_z E_z \end{aligned} \quad (S3.4)$$

The analytically calculated force densities are shown in Fig. S3. When the tip is on top of a single domain (Fig. S3A), $F_x$ is an odd function (antisymmetric) with respect to $x = x_{\text{tip}}$, whereas $F_y$ and $F_z$ are even functions (symmetric) with respect to the tip. On the other hand, due to the sign flip of the piezoelectric tensor in opposite domains, the parity of all three components of the force density is reversed when the tip is on top of the DW (Fig. S3B). As discussed in the main text, if we only consider the y-component of the P-SAW, the time derivative of the displacement field $\dot{u}_y$ is always an even function with respect to $x = x_{\text{tip}}$ due to the continuity condition. Hence the overlap integral between $F_y$ and $\dot{u}_y$ cancels each other in the two shaded areas in Fig. 2A.

**Section S4. Details of the FEA modeling.**

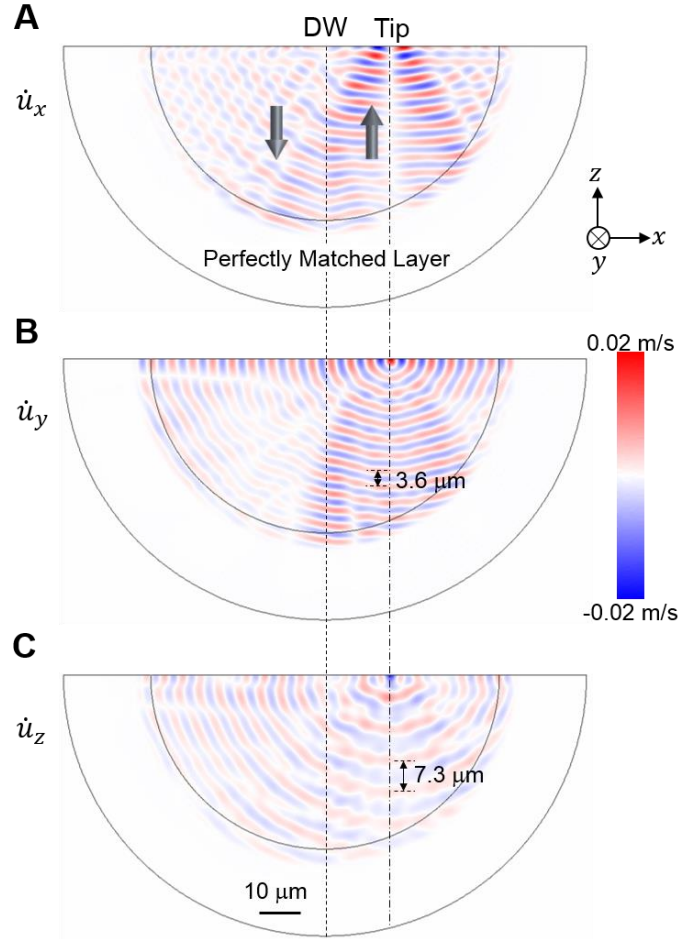

**Fig. S4**. Snap shots of (*A*) *x*-, (*B*) *y*-, and (*C*) *z*-components of the simulated velocity (time derivative of the displacement) fields. The inner half-circle is LiNbO$_3$ and the outer half-ring is the perfectly matched layer (PML). The DW (dashed line) lies in the middle of the sample. The slow transverse and longitudinal bulk waves propagating along the *z*-direction can be seen from the $\dot{u}_y$ and $\dot{u}_z$ plots, respectively.

The tip-excited acoustic waves in LiNbO$_3$ can be numerically simulated by finite-element analysis (FEA). Since a full 3D FEA requires excessive computer memory, we simulate a thin plate (1 μm in thickness) with periodic boundary condition along the *y*-direction. The finite thickness is to enable the motion in the *y*-axis. As stated in the main text, such an approximation captures the essential physics in the MIM experiment, except that the tip-induced *E*-field decreases as $1/r$ from the tip rather than $1/r^2$ in the actual case. The DW separating the two oppositely polarized domains is positioned in the *yz*-plane. The LiNbO$_3$ region, whose permittivity, piezoelectric coefficient, and elasticity tensor are taken from the literature [*S6*, *S7*], is bounded by the perfectly matched layer (PML) to avoid wave reflection from the boundary. The tip on the sample surface is modeled as a line source along the *y*-axis with an oscillating voltage of $V = V_0 e^{i2\pi ft}$, where $V_0 = 1$ V and $f = 1$ GHz. With these input conditions, the FEA software (COMSOL4.3, Structural Mechanics module, linear solver) can compute the time-dependent displacement ($\vec{u}$) fields, from which the velocity ($\dot{\vec{u}}$), electric field ($\vec{E}$), current density ($\vec{j}$), and other physical parameters can be derived and analyzed.

Fig. S4 shows snap shots ($t = 0$) of $\dot{\vec{u}}$ fields when the tip is 15 μm away from the DW. A video clip showing the time evolution of the propagating waves is also included in the Supplementary Information Movie S1. As is evident from the plots, the tip excites several acoustic modes due to the piezoelectric coupling. Near the surface, the pseudo surface acoustic wave (P-SAW) is clearly seen in the $\dot{u}_y$ map since the P-SAW is mostly transverse-horizontal. The Rayleigh-type SAW, on the other hand, is predominately polarized along the *z*-axis. Consequently, the beat between SAW ($\lambda = 3.82$ μm at 1 GHz) and P-SAW ($\lambda = 4.55$ μm at 1 GHz) appears in the $\dot{u}_z$ map. The leaking of energy from the P-SAW to the slow transverse bulk wave is obvious in all three plots. Finally, for the bulk waves propagating along the *z*-axis, the slow transverse wave ($\lambda = 3.6$ μm at 1 GHz) and the longitudinal wave ($\lambda = 7.3$ μm at 1 GHz) can be seen from the $\dot{u}_y$ and $\dot{u}_z$ plots, respectively. Note that these bulk velocities [S8] are different from the values in Table 1, which are along the *x*-axis of LiNbO$_3$.

## Section S5. P-SAW versus SAW

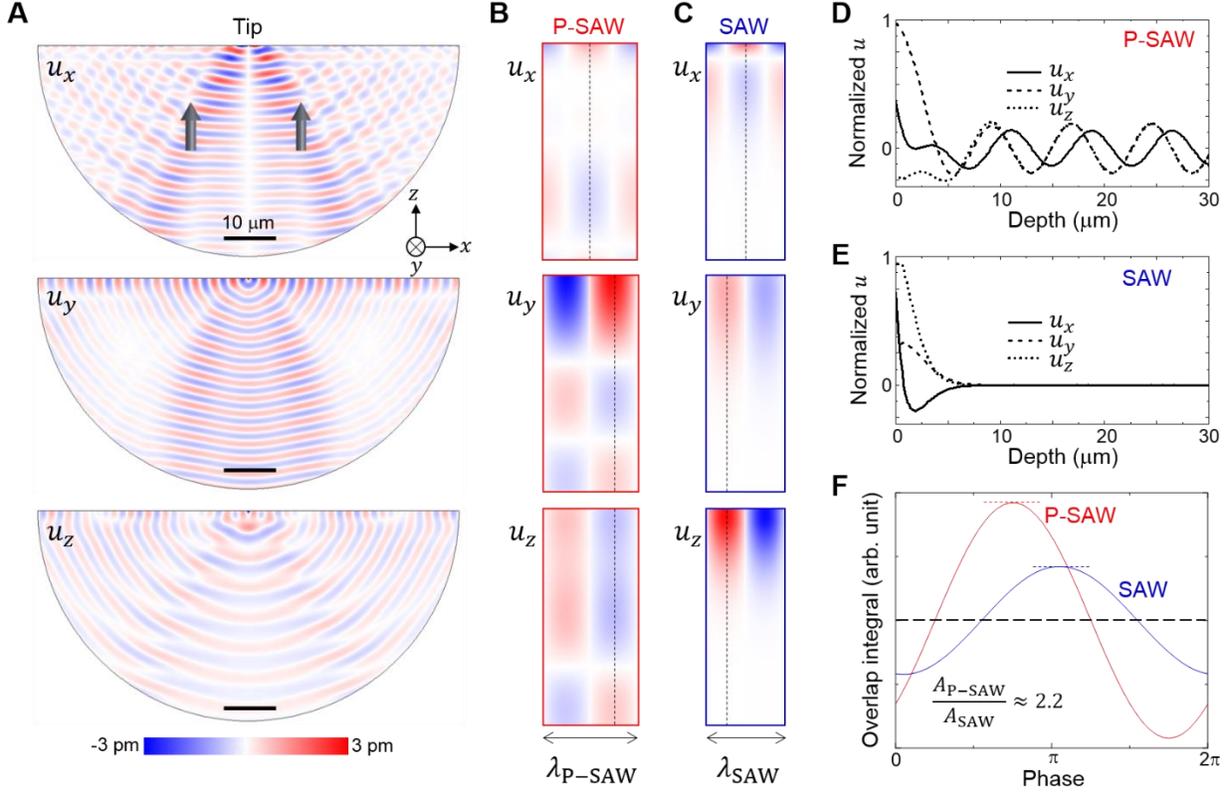

**Fig. S5**. (*A*) (Top to bottom) Snap shots of *x*-, *y*-, and *z*-components of the simulated displacement fields when the tip is on top of a single domain. The PML region is not shown. (*B*) Eigen modes of the P-SAW and (*C*) SAW displacement fields. Note that the two waves have different wavelengths at 1 GHz due to the different velocities. (*D*) Depth profiles along the dashed lines in (*B*) and (*C*) of the normalized P-SAW and (*E*) SAW displacement fields. (*F*) Overlap integral between the linear-solver results in (*A*) and eigen-solver results in (*B*) and (*C*). The relative strength between P-SAW and SAW excited by the tip voltage can be extracted.

The tip voltage at microwave frequencies is capable of exciting both the SAW and P-SAW. In order to analyze the relative strength between the two waves, we first use the linear solver to compute the displacement fields when the tip is on top of a single domain, as shown in Fig. S5A. The beat between P-SAW and SAW can be seen near the surface in the $u_z$ plot. Next, we solve the eigen-modes of P-SAW $\vec{u}_{\text{P-SAW}}$ (Fig. S5B) and SAW $\vec{u}_{\text{SAW}}$ (Fig. S5C) by FEA. The leaky nature of the P-SAW (displacement not confined to the surface) is evident in the depth profile in Fig. S5D. Note that the P-SAW is primarily polarized in the *y*-direction on the surface, reminiscent of the electroacoustic Bleustein-Gulyaev SAW [*S2*, *S3*]. In contrast, the non-leaky SAW is confined to the surface and its primary polarization is along the *z*-axis.

The tip-induced displacement $\vec{u}$ in Fig. S5A contains both P-SAW and SAW on the surface. To analyze their relative strength, we take the overlap integral of $\int \vec{u} \cdot \vec{u}_{\text{P-SAW}} dx$ and $\int \vec{u} \cdot \vec{u}_{\text{SAW}} dx$ within the dashed box in Fig. S5A, where the phase of $\vec{u}$ varies continuously from 0 to $2\pi$. The result in Fig. S5F shows that the amplitude of P-SAW is ~ 2.2 times of the SAW. In other words, the transduction from electrical energy into mechanical energy is ~ 5 times more effective for the P-SAW than that for the SAW, which explains the dominance of P-SAW in the experiment. Note that since there are only a few oscillations in the MIM-Re data (Fig. 1C), it is difficult to separate SAW ($\lambda$ = 3.82 μm at 1 GHz) and P-SAW ($\lambda$ = 4.55 μm at 1 GHz) from the Fourier transform.

## Section S6. Complete data of the enclosed domains

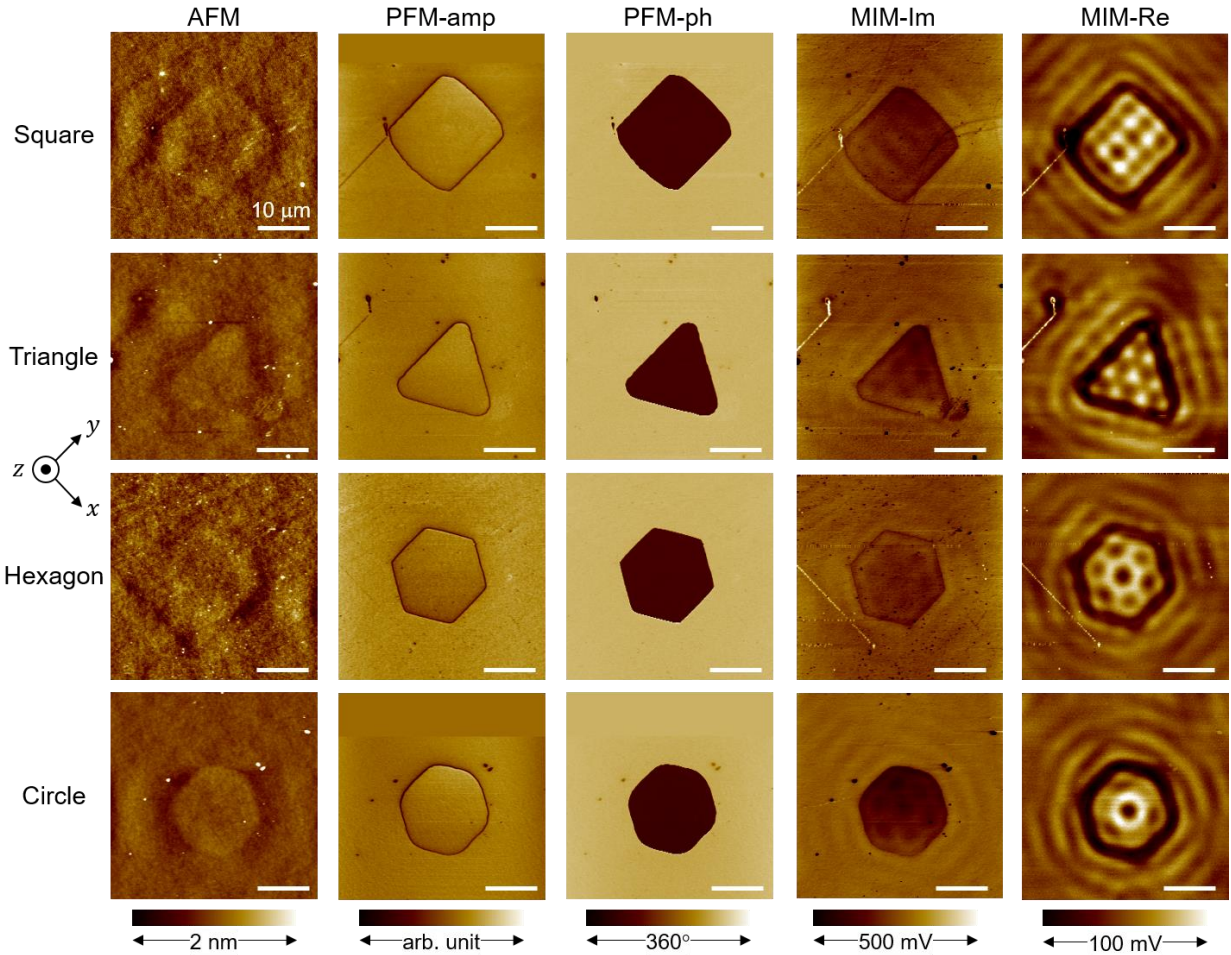

**Fig. S6**. Complete set of data of the enclosed domains. All scale bars are 10 μm.

The complete set of data for Fig. 4 in the main text is shown in Fig. S6. The sample surface is very smooth after the electrical poling. As expected, the opposite $LiNbO_3$ domains display DW contrast in the PFM amplitude image and 180° contrast in the PFM phase image. The MIM-Im images exhibit weak contrast since the tip-sample susceptance is dominated by the dielectric response. The interference patterns in the MIM-Re data are due to piezoelectric transduction, which is analyzed in the main text.